\documentclass[runningheads,a4paper]{llncs}

\usepackage{amssymb}
\usepackage{graphicx}
\usepackage{booktabs}
\usepackage{xcolor}
\usepackage{multirow}
\usepackage{amsmath}

\usepackage{url}
\newcommand{\keywords}[1]{\par\addvspace\baselineskip
\noindent\keywordname\enspace\ignorespaces#1}

\begin{document}

\mainmatter  %

\title{Image Classification via Quantum Machine Learning}

\titlerunning{QML: Image Classification}

\author{Héctor Iván García-Hernández%
\and Raymundo Torres-Ruiz\and Guo-Hua Sun%
}
\authorrunning{Image Classification via Quantum Machine Learning}

\institute{Instituto Politécnico Nacional, Centro de Investigación en Computación, Mexico City, Mexico \\
}

\toctitle{Quantum Machine Learning}
\tocauthor{Image Classification}
\maketitle

\begin{abstract}
Quantum Computing and especially Quantum Machine Learning, in a short period of time, has gained a lot of interest through research groups around the world. This can be seen in the increasing number of proposed models for pattern classification applying quantum principles to a certain degree. Despise the increasing volume of models, there is a void in testing these models on real datasets and not only on synthetic ones. The objective of this work is to classify patterns with binary attributes using a quantum classifier. Specially, we show results of a complete quantum classifier applied to image datasets. The experiments show favorable output while dealing with balanced classification problems as well as with imbalanced classes where the minority class is the most relevant. This is promising in medical areas, where usually the important class is also the minority class.
\end{abstract}

\section{Introduction}

Image classification is of utmost importance in several areas of science and technology such as medical diagnosis and prognosis, face detection, or multiple object detection for autonomous cars. By classical models, this task can be solved using Convolutional Neural Networks~\cite{LeCun1998} but it is notorious the enormous number of parameters needed to train, as seen, for example, in~\cite{Kaiming2016}. Nevertheless, exploiting the prowess of Quantum Mechanics such as interference, superposition, and entanglement, which promises great power of computation and in compass with the recent implementation of several quantum computers, it is worth to propose and evaluate quantum models for machine learning. Although these models are, in essence, simple and with performances lower than the state of the art, they serve as stepping stones for the construction of increasingly complex models with much better performance.

The \textbf{advantage} of the quantum models is the inherent parallelism in their execution, the speed at which they are executed, and even more important is the exponential reduction of the number of qubits necessary to encode the information compared to the classical models, for example, only \(6\) qubits are needed to encode a \(64\)-dimensional pattern and with just \(30\) qubits we could encode a \(32768\) x \(32768\) binary image, which is more than a billion-dimensional flatten vector. This reduction is possible by exploiting superposition states and quantum entanglement.

Let us present some high-level descriptions of models proposed by various research groups, either purely quantum or hybrid combining classical and quantum processing. Yamamoto et al.~\cite{Yamamoto2018} proposed a quantum perceptron model that allows classifying non-linearly separable data. Maria Schuld et al.~\cite{Schuld2015} proposed another quantum perceptron model using unitary operators acting on two qubits and the inverse quantum Fourier transform. Maxwell Henderson et al.~\cite{henderson2019} put forward a quantum convolutional layer model for the extraction of features in images. Sebastien Piat et al.~\cite{Piat2018} proposed a preprocessing with auto-encoders, a restricted Boltzmann machine (RBM) is trained in a quantum computer. This RBM is used to initialize a classical neural network which is subsequently trained in a classical way.
Iris Cong et al.~\cite{Cong_2019} utilized another model for a quantum convolutional network. Zhao et al.~\cite{Zhao2019} proposed a swap-based red neural quantum test. Dang et al.~\cite{Dang2018} proposed a KNN-based quantum classifier, with a classical model for feature extraction. Francesco et al.~\cite{Francesco2019} proposed a new model for a quantum neuron implemented in a real quantum processor. Using Qiskit~\cite{Qiskit} and Pytorch, arbitrarily large hybrid models can be generated~\cite{qiskitpytorch}. Despite having various proposals and with various applications~\cite{Jeswal2018}, each of the models omits a feasible implementation in a real quantum processor, they lack a proof in a real dataset or in their most extreme case they do not correctly use quantum mechanics~\cite{Zhou1999}.

	With this work, we aim to show the potential application of Quantum Machine Learning to real-world problems in image classification validating and testing 
a quantum classification model beyond the theoretical realm and the standard proof of concept.

	In this work we first explore two real image datasets, some performance measures are discussed, and the implementation of a quantum classifier is described both at a theoretical and at a high level in Section~\ref{section:section3}. The performance of the classifier in the two datasets is presented in Section~\ref{section:section4}, evaluating its performance when facing a problem of both, balanced classes and highly unbalanced classes. Finally, the conclusions and future work are presented in Section~\ref{section:section5}.

\section{Datasets and Algorithms}
\label{section:section3}

In this section, we summarize two digits images datasets. The performance measures used to evaluate the classifier are discussed and the classifier itself is analyzed. Before starting to introduce them, it is necessary to propose the theoretical description to be used in this work.

\subsection{Theoretical Model}

Prior to describing the practical steps in the algorithm, a theoretical approach must be taken, so with a little more in-depth description, and following~\cite{Francesco2019} almost verbatim, we start with the binary pattern \(\vec{x}^{T} = \left(x_{0} \ldots x_{m-1}\right)\) and the weight vector \(\vec{\Omega}^{T} = \left(\Omega_{0} \ldots \Omega_{m-1}\right)\) with \(x_{j}, \Omega_{j} \in\{0, 1\}\) and then we map them to

\begin{equation}
\label{eqn:eq5}
	\vec{i}=\left(\begin{array}{c}
	i_{0} \\
	i_{1} \\
	\vdots \\
	i_{m-1}
	\end{array}\right), \vec{w}=\left(\begin{array}{c}
	w_{0} \\
	w_{1} \\
	\vdots \\
	w_{m-1}
	\end{array}\right)
\end{equation}

with  \(i_{j}, w_{j} \in\{-1, 1\}\) and with them we can define two quantum states

\begin{equation}
\label{eqn:eq6}
\left|\psi_{i}\right\rangle=\frac{1}{\sqrt{m}} \sum_{j=0}^{m-1} i_{j}|j\rangle \text{	and	} \left|\psi_{w}\right\rangle=\frac{1}{\sqrt{m}} \sum_{j=0}^{m-1} w_{j}|j\rangle\; .
\end{equation}

The states \(|j\rangle \in\{|00 \ldots 00\rangle,|00 \ldots 01\rangle, \ldots,|11 \ldots 11\rangle\}\) form the computational basis of a quantum processor.
If \(N\) qubits are used in the register, there are \(m=2^{N}\) basis states labeled \(|j\rangle\) and we can use factors \(\pm 1\) to encode the \(m\) -dimensional classical patterns and weights into a uniformly weighted superposition of the computational basis.

The first step is to prepare the state \(\left|\psi_{i}\right\rangle\) by encoding the input values of \(\vec{i}\). With the qubits initialized in the zero state \(|00 \ldots 00\rangle \equiv|0\rangle^{\otimes N},\) we perform a unitary transformation \(U_{i}\)

\begin{equation}
\label{eqn:eq7}
U_{i}|0\rangle^{\otimes N}=\left|\psi_{i}\right\rangle\; .
\end{equation}

The second step computes the inner product between \(\vec{w}\) and \(\vec{i}\) using the quantum register. This can be done defining a unitary transformation, \(U_{w}\), such that

\begin{equation}
\label{eqn:eq8}
U_{w}\left|\psi_{w}\right\rangle=|1\rangle^{\otimes N}=|m-1\rangle\; .
\end{equation}

If we apply \(U_{w}\) after \(U_{i}\), the quantum state becomes

\begin{equation}
\label{eqn:eq9}
U_{w}\left|\psi_{i}\right\rangle=\sum_{j=0}^{m-1} c_{j}|j\rangle \equiv\left|\phi_{i, w}\right\rangle\; .
\end{equation}

Using Eq.~(\ref{eqn:eq8}), the scalar product between the two quantum states is

\begin{equation}\begin{aligned}
\label{eqn:eq10}
\left\langle\psi_{w} \mid \psi_{i}\right\rangle &=\left\langle\psi_{w}\left|U_{w}^{\dagger} U_{w}\right| \psi_{i}\right\rangle \\
&=\left\langle m-1 \mid \phi_{i, w}\right\rangle=c_{m-1}\; ,
\end{aligned}\end{equation}
and from the definitions in Eq.~(\ref{eqn:eq6}) we see that the scalar product of input and weight vectors is \(\vec{w} \cdot \vec{i}=m\left\langle\psi_{w} \mid \psi_{i}\right\rangle\). Hence, the desired result is contained, up to a normalization factor, in the coefficient \(c_{m-1}\) of the final state \(\left|\phi_{i, w}\right\rangle\).

In order to extract such an information, an ancilla qubit \((a)\) initially set in the state \(|0\rangle\) is toggled by a multi-controlled NOT gate between the \(N\) encoding qubits, this leads to

\begin{equation}
\label{eqn:eq11}
\left|\phi_{i, w}\right\rangle|0\rangle_{a} \rightarrow \sum_{j=0}^{m-2} c_{j}|j\rangle|0\rangle_{a}+c_{m-1}|m-1\rangle|1\rangle_{a}\; .
\end{equation}

The nonlinearity required by the threshold function at the output of the perceptron is immediately obtained by performing a quantum measurement. By measuring the state of the ancilla qubit produces the output \(|1\rangle_{a}\) with probability \(\left|c_{m-1}\right|^{2}\).

	Even though general advantages or disadvantages cannot be outlined, we can mention some of the differences between this quantum model and a classical one: an execution of a classical model gives an activation directly comparable to a threshold whereas the quantum model gives a binary output which must be repeated several times in order to acquire a measure approximately to theory and comparable to the threshold. Another difference is that since the quantum measure is binary, it can be readily interpreted as the assigned class given the nonlinearity function is already satisfied. And lastly, in general, a classical model does not require input patterns with numerical attributes to be codified into other representations, although it can be useful, however, most quantum models, including this one,  are limited to patterns with binary attributes or to be coded as such.

\subsection{Datasets}

Two databases are used for the experiments. They are both images of digits. The first is the “Optical Recognition of Handwritten Digits Data Set (Digits Dataset)”, which contains 64 attributes in a range from 0 to 16 with 5620 total instances~\cite{Alpaydin1997}. The test set with 1797 instances, is available directly on the scikit-learn Python package. The second database is the "Semeion Handwritten Digit Data Set (Semeion Dataset)" which contains 1593 instances each with 256 binary attributes~\cite{UCI}. Both datasets are balanced, containing approximately the same number of instances per class.

\begin{figure}
\centering
\includegraphics[height=6.2cm]{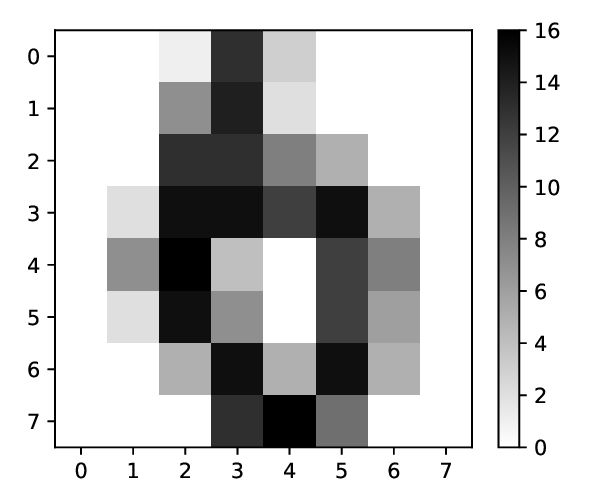}
\caption{An instance for class \(6\) in the Digits Dataset. Each pixel has a value varying from \(0\) up to \(16\). As can be seen, the low resolution can lead to misclassification, since this instance can be easily mistaken for an instance of class \(0\).}
\label{fig:example}
\end{figure}

\begin{figure}
\centering
\includegraphics[height=6.2cm]{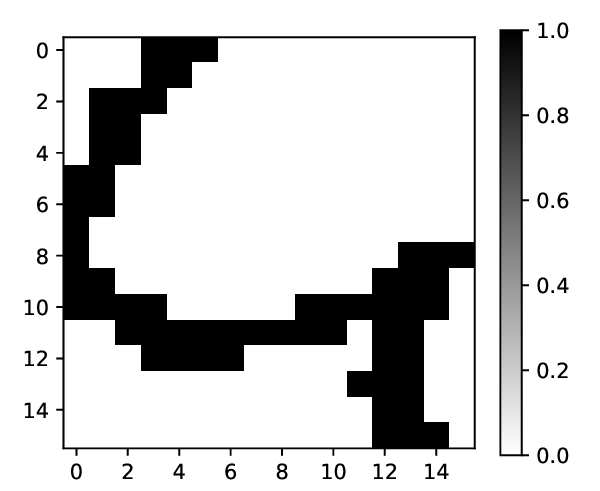}
\caption{An instance for class \(4\) in the Semeion Dataset. Each attribute, or pixel, has a binary value. These patterns have a bigger dimensionality, and thus are less likely to be misclassified, but also the complexity of the processing needed to build the quantum circuit is increased.}
\label{fig:example}
\end{figure}

\begin{table}[]
\caption{Summary of both used digits images datasets.}
\label{tab:datasets}
\centering
\begin{tabular}{@{}cllcllcllc@{}}
\toprule
Dataset &  &  & Classes &  &  & Imbalance Ratio &  &  & Total Instances \\ \midrule
Digits  &  &  & 10      &  &  & 1.032           &  &  & 5620            \\
Semeion &  &  & 10      &  &  & 1.045           &  &  & 1593            \\ \bottomrule
\end{tabular}
\end{table}

\subsection{Performance Measures}

When each class contains roughly the same number of instances in a dataset, it is known as a balanced dataset. In these cases, most of the performance measures are adequate, as long as there is no bias towards any class. However, depending on the application of the classifier and the relevance of any of the classes, some other performance measure may be chosen.

\begin{table}[]
\caption{Confusion matrix for a two-class dataset.}
\label{tab:confusion-matrix}
\centering
\begin{tabular}{@{}clclclc@{}}
\cmidrule(l){5-7}
                           &  &          &  & \multicolumn{3}{c}{Real   Class}                 \\ \cmidrule(l){5-7} 
                           &  &          &  & Positive              &  & Negative              \\ \cmidrule(l){2-7}
\multirow{2}{*}{Predicted Class} &  & Positive &  & True   Positive (TP)  &  & False   Positive (FP) \\ \cmidrule(l){2-7}
                           &  & Negative &  & False   Negative (FN) &  & True   Negative (TN)  \\ \cmidrule(l){2-7}
\end{tabular}
\end{table}

The most common is accuracy, which measures the ratio of instances correctly classified to the total number of instances.

\begin{equation}
\label{eqn:acc}
  accuracy = \frac{TP + TN}{TP + FP + FN + TN}\;  .
\end{equation}

	A dataset is unbalanced when one or more classes is poorly represented in the dataset. Most classical performance measures produce a majority class bias in an unbalanced class problem. In these cases the True Positive Rate (TPR)

\begin{equation}
\label{eqn:tpr}
  TPR = \frac{TP}{TP + FN}\;  .
\end{equation}

	which is also known as Recall or Sensitivity can be used to measure the ratio of the number of positive instances correctly classified to the total number of positive instances.

	We also keep track of the Positive Predictive Value (PPV)(also known as Precision)

\begin{equation}
\label{eqn:ppv}
  PPV = \frac{TP}{TP + FP}\;  .
\end{equation}

	which measures the ratio of the number of positive instances correctly classified to the total number of positive classified instances. With TPR and PPV the $F1$ score can be obtained, which is the harmonic mean of these performance measures.

\begin{equation}
\label{eqn:f1}
  F1 = 2*\frac{PPV * TPR}{PPV + TPR} = \frac{2 * TP}{2 * TP + FP + FN}\;  .
\end{equation}

When classes contain insufficient instances to be partitioned in a traditional validation method, it is common to use all instances for training and testing process. Accuracy under this validation method is known as Resubstitution Error.

\subsection{Quantum Classification Algorithm}

The model used for the classification task is an implementation of the one described in~\cite{Francesco2019}. In this model, an instance with binary attributes is encoded by means of a method called \emph{hypergraph states generation subroutine}~\cite{Francesco2019,Rossi2013}. The weight vector is randomly initialized, which will be updated accordingly to a set of hyper-parameters, which will regulate its rate of change. At the end of the execution of the dynamically generated quantum circuit through the process described in~\cite{Francesco2019}, a measurement is performed on the ancilla qubit, which will take the value \(0\) or \(1\). By means of several repetitions of the circuit, the proportion of measurements with result \(1\) over the total measurements can be obtained. The more measures are made, the closer the result is to the real one.

\begin{figure}
\centering
\includegraphics[height=6.6cm]{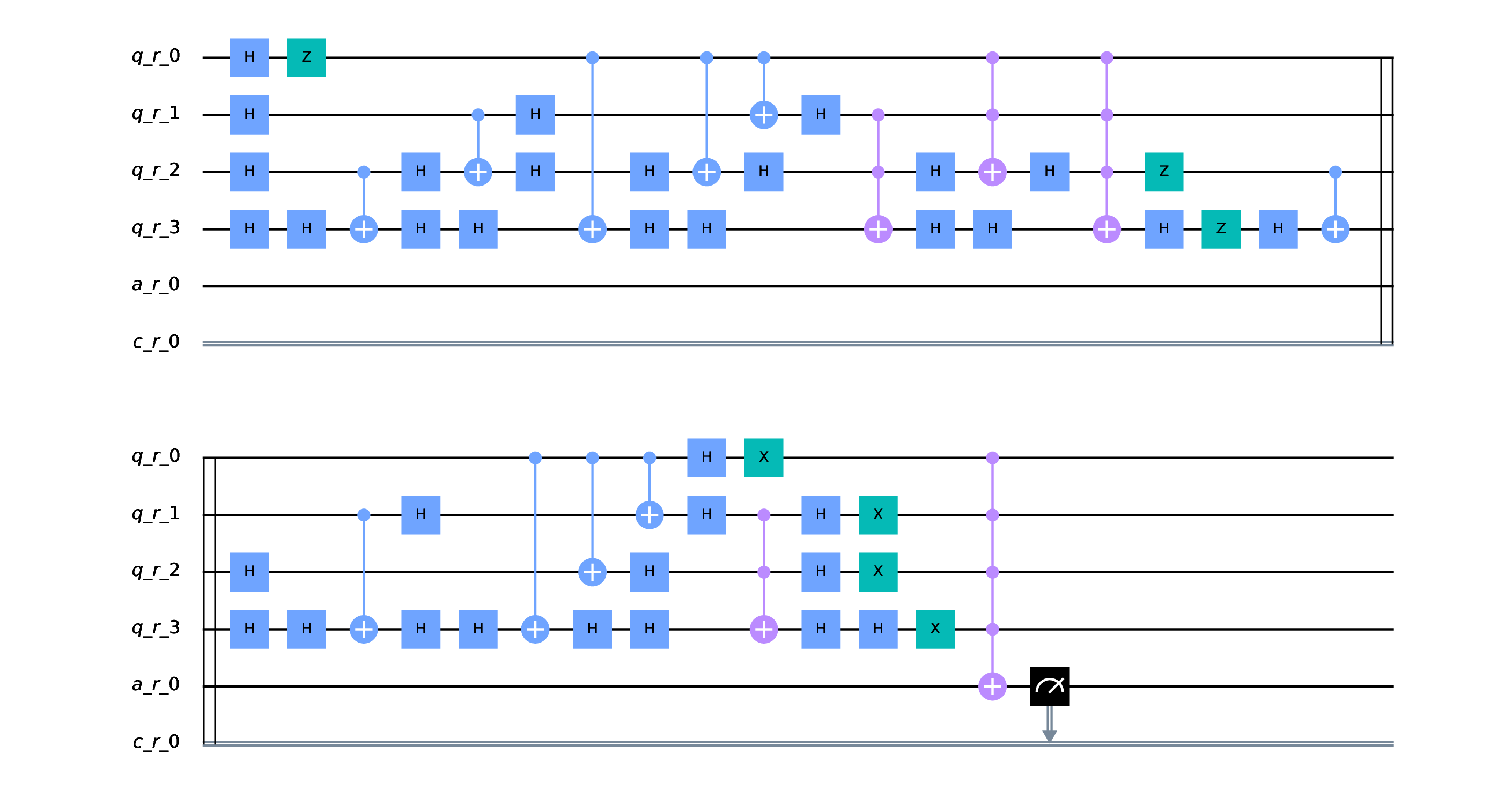}
\caption{Quantum circuit programmatically generated encoding both a pattern and a vector weight. The circuit is encoding an image of size \(4\)x\(4\) using \(4\) qubits for processing, one qubit as ancilla, and one classical bit storing the measurement. This relatively simple circuit already contains \(34\) layers of quantum gates, this can convey the magnitude of the circuits needed to process the \(16\)x\(16\) images. Currently bounded by the physical implementation of real quantum computers, the model is not limited by the pattern dimensions it can process.}
\label{fig:example}
\end{figure}

This proportion, which we called \emph{readout}, is compared with another hyper-parameter, which is called threshold. This threshold is used to assign the class. If the readout is less than the threshold the positive class is assigned, otherwise the negative class is assigned to the pattern in turn.

\begin{figure}
\centering
\includegraphics[height=6.6cm]{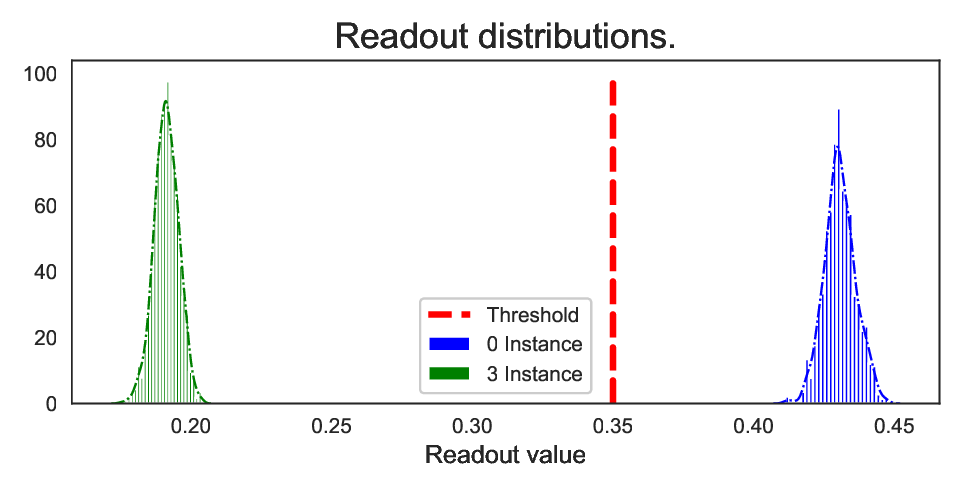}
\caption{Readouts distributions after a thousand iterations of the same circuit using the final weight vector for the \(3\slash 0\) binary classifier. The distribution at the left is for an instance of the class \(3\) and is well below the threshold. The distribution at the right is for an instance of class \(0\), this time, above the threshold.}
\label{fig:example}
\end{figure}

As it is a supervised classification task, during the training step, it will be evaluated if the assigned class is correct. If it is, we simply continue with the next pattern. In the case it is incorrectly classified, depending on its real class, a hyper-parameter will be used, in a similar fashion to the traditional learning rate in neural networks, which defines the proportion of change in the weight vector. There is a learning rate for the positive class and another one for the negative class.

As usual, this procedure can be repeated for an arbitrary number of epochs, where one epoch means that the classifier has seen the entire training set. Or also, as it was done, the training can be finished earlier if a critical value has been reached in a certain metric that we seek to optimize.

The whole process, aiming for efficiency, is simulated using Qiskit~\cite{Qiskit}, however, the entire process is suitable and ready to be executed on a real quantum machine.

	Although we implemented and followed the model described by~\cite{Francesco2019}, we diverge from them in the sense that we do not test the model in an \emph{ad hoc} dataset. Furthermore, the \emph{ad hoc} dataset was split by means of a previously and arbitrarily  selected weight vector. In other words, the classification task was already known to be solvable because it was constructed to do so, but in this work, we do not assume or fabricate such property and let the model converge to a solution.

\section{Experimental Results and Discussion}
\label{section:section4}

In this section we present the experimental results of the classification model described above. The model was tested in both Optical Recognition of Handwritten Digits Data Set and Semeion Handwritten Digit Data Set.

In the case of Digits, Resubstitution Error and Hold-out were used as validation method. For Hold-out, we used the partition offered by the authors~\cite{Alpaydin1997}. For Resubstitution Error, only the test set was used, acting as both training and test sets. This dataset requires processing, as each attribute has a value between \(0\) and \(16\), and the model only works with binary values a threshold was applied to binarize the patterns as follows:

\begin{equation}
	\text{binarized pixel} =
		\begin{cases}
			0,		& \text{if original pixel value} < 10 \\
			1,		& \text{otherwise}
		\end{cases}
\end{equation}

\begin{figure}
\centering
\includegraphics[height=6.2cm]{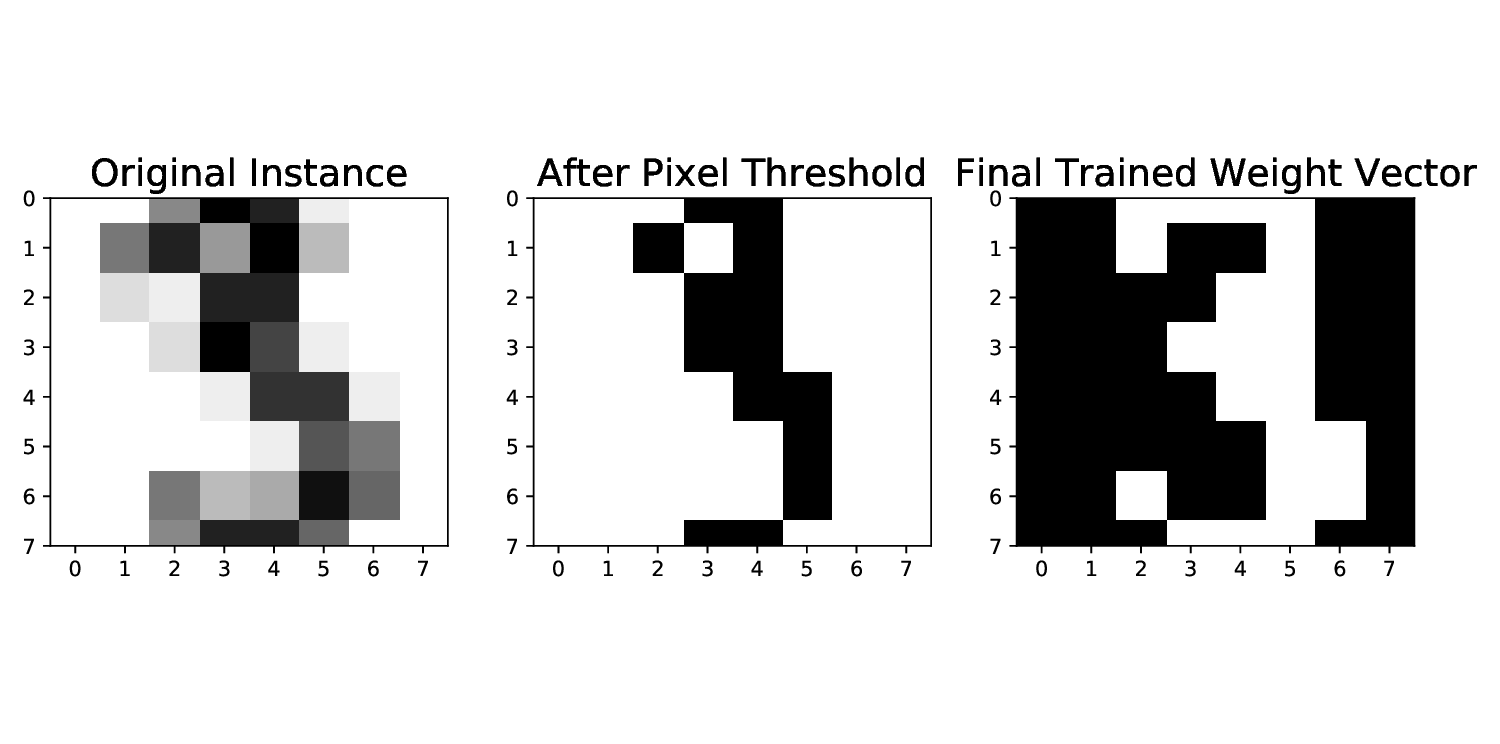}
\caption{At the left is an original instance from the Digits Dataset. In the middle is the result of the binarization threshold. And at the right is the final weight vector for the \(0\slash 3\) classifier. The vector tends to take the form of sort of a mask for one of the classes.}
\label{fig:example}
\end{figure}

This threshold is itself a new hyper-parameter that can be optimized.

In the case of Semeion, Resubstitution Error was used due to the relatively small number of available patterns. This dataset does not need processing as the attributes are already binary.

In each dataset, binary classifiers of two styles are trained and tested: class vs class, also known as One vs One (OvO), which in this case represents a balanced classes problem, and class vs the rest also known as One vs All (OvA) which represents an unbalanced class problem. The results are shown in several tables.
In Table~\ref{tab:digits-ovo-res} the diagonal represents the trivial classification of one single class. Is interesting to note although we can use the upper or lower half to classify the reflexive class, i.e. use the trained positive/negative classifier to try to classify the negative/positive problem, we did not get good results in this scenario. This makes sense when the vector weight for each binary classifier is inspected since it tends to take the form of a sort of mask resembling the instances of the negative class. In Table~\ref{tab:digits-ova-res} we keep track of some performance measures, including the Area Under the Curve (AUC). It is evident from the Recall measure that the quantum model can distinguish the positive class from all the other instances with good accuracy. The ratio of minority class against the rest is approximately 1:100. This result is promising for medical applications where datasets are generally heavily imbalanced. We show, in Table~\ref{tab:digits-ho-ovo} the increase in accuracy performance when a different validation method is used. It is also notable the latent power of generalization because in this case the evaluation was recorded upon a test set containing instances not seen during the training set. In Table~\ref{tab:digits-ho-ova} we show the usual metrics, where is notorious the benefit gained by the Hold-out method for this classification model in this dataset. We noted an improvement in seven out of the ten classes. This validation method gives us more confidence in the generalization power that this model might have. In Table~\ref{tab:seme-ovo-res} and in Table~\ref{tab:seme-ova-res} we give the already known performance metrics. Acceptable performance can be seen in almost every classification task, but it is notorious the decrease compared to the Digits dataset. This drop can be explained by the fact that each instance lives in a bigger dimensional space, and therefore the quantum circuit needed to process each pattern is also bigger and more complex quantum-gates-wise. Nevertheless, we must keep in mind although both datasets might seem similar they are in fact different and we should not expect the same performance in both of them.

\begin{table}[]
\caption{Resubstitution Error by balanced class classification.}
\label{tab:digits-ovo-res}
\centering
\begin{tabular}{@{}ccccccccccccccccccccccccccccc@{}}
\toprule
\multicolumn{29}{c}{Digits Resubstitution Error}                                                                                                    \\ \midrule
                \multicolumn{28}{c}{Negative Class}                                                                                                 \\ \midrule
               & 0     &  &  & 1     &  &  & 2     &  &  & 3     &  &  & 4     &  &  & 5     &  &  & 6     &  &  & 7     &  &  & 8     &  &  & 9     \\ \midrule
Positive Class &       &  &  &       &  &  &       &  &  &       &  &  &       &  &  &       &  &  &       &  &  &       &  &  &       &  &  &       \\ \midrule
0              & 1.0   &  &  & 0.725 &  &  & 0.794 &  &  & 0.714 &  &  & 0.671 &  &  & 0.769 &  &  & 0.353 &  &  & 0.834 &  &  & 0.678 &  &  & 0.589 \\ \midrule
1              & 0.377 &  &  & 1.0   &  &  & 0.660 &  &  & 0.849 &  &  & 0.749 &  &  & 0.780 &  &  & 0.898 &  &  & 0.833 &  &  & 0.609 &  &  & 0.759 \\ \midrule
2              & 0.833 &  &  & 0.710 &  &  & 1.0   &  &  & 0.730 &  &  & 0.804 &  &  & 0.894 &  &  & 0.631 &  &  & 0.764 &  &  & 0.564 &  &  & 0.820 \\ \midrule
3              & 0.933 &  &  & 0.753 &  &  & 0.833 &  &  & 1.0   &  &  & 0.780 &  &  & 0.887 &  &  & 0.807 &  &  & 0.676 &  &  & 0.624 &  &  & 0.652 \\ \midrule
4              & 0.974 &  &  & 0.567 &  &  & 0.837 &  &  & 0.758 &  &  & 1.0   &  &  & 0.785 &  &  & 0.792 &  &  & 0.780 &  &  & 0.749 &  &  & 0.764 \\ \midrule
5              & 0.886 &  &  & 0.785 &  &  & 0.830 &  &  & 0.715 &  &  & 0.779 &  &  & 1.0   &  &  & 0.953 &  &  & 0.695 &  &  & 0.800 &  &  & 0.640 \\ \midrule
6              & 0.635 &  &  & 0.807 &  &  & 0.843 &  &  & 0.807 &  &  & 0.682 &  &  & 0.785 &  &  & 1.0   &  &  & 0.883 &  &  & 0.540 &  &  & 0.828 \\ \midrule
7              & 0.983 &  &  & 0.653 &  &  & 0.803 &  &  & 0.776 &  &  & 0.875 &  &  & 0.797 &  &  & 0.905 &  &  & 1.0   &  &  & 0.728 &  &  & 0.832 \\ \midrule
8              & 0.752 &  &  & 0.480 &  &  & 0.706 &  &  & 0.792 &  &  & 0.783 &  &  & 0.651 &  &  & 0.830 &  &  & 0.597 &  &  & 1.0   &  &  & 0.666 \\ \midrule
9              & 0.564 &  &  & 0.812 &  &  & 0.815 &  &  & 0.696 &  &  & 0.883 &  &  & 0.527 &  &  & 0.678 &  &  & 0.813 &  &  & 0.788 &  &  & 1.0   \\ \bottomrule
\end{tabular}
\end{table}

\begin{table}[]
\caption{Resubstitution Error by heavily imbalanced class classification.}
\label{tab:digits-ova-res}
\centering
\begin{tabular}{@{}cccccccccc@{}}
\toprule
\multicolumn{10}{c}{Digits Resubstitution Error}                         \\ \midrule
Positive Class & Recall &  & Accuracy &  & Precision &  & F1    &  & AUC   \\ \midrule
0              & 0.983  &  & 0.303    &  & 0.122     &  & 0.218 &  & 0.605 \\ \midrule
1              & 0.857  &  & 0.399    &  & 0.129     &  & 0.224 &  & 0.602 \\ \midrule
2              & 0.966  &  & 0.377    &  & 0.133     &  & 0.234 &  & 0.639 \\ \midrule
3              & 0.814  &  & 0.328    &  & 0.112     &  & 0.198 &  & 0.543 \\ \midrule
4              & 0.955  &  & 0.420    &  & 0.143     &  & 0.249 &  & 0.657 \\ \midrule
5              & 0.873  &  & 0.359    &  & 0.123     &  & 0.216 &  & 0.587 \\ \midrule
6              & 0.845  &  & 0.368    &  & 0.121     &  & 0.212 &  & 0.580 \\ \midrule
7              & 0.837  &  & 0.417    &  & 0.128     &  & 0.222 &  & 0.604 \\ \midrule
8              & 0.724  &  & 0.388    &  & 0.107     &  & 0.186 &  & 0.538 \\ \midrule
9              & 0.861  &  & 0.419    &  & 0.132     &  & 0.229 &  & 0.615 \\ \bottomrule
\end{tabular}
\end{table}

\begin{table}[]
\caption{Accuracy measure by balanced class classification.}
\label{tab:digits-ho-ovo}
\centering
\begin{tabular}{@{}ccccccccccccccccccccccccccccc@{}}
\toprule
\multicolumn{29}{c}{Digits   Hold-out}                                                                                                               \\ \midrule
\multicolumn{29}{c}{Negative Class}                                                                                                                  \\ \midrule
               & 0     &  &  & 1     &  &  & 2     &  &  & 3     &  &  & 4     &  &  & 5     &  &  & 6     &  &  & 7     &  &  & 8     &  &  & 9     \\ \midrule
Positive Class & \multicolumn{28}{c}{}                                                                                                               \\ \midrule
0              & 1.0   &  &  & 0.844 &  &  & 0.833 &  &  & 0.900 &  &  & 0.896 &  &  & 0.833 &  &  & 0.924 &  &  & 0.831 &  &  & 0.732 &  &  & 0.731 \\ \midrule
1              & 0.969 &  &  & 1.0   &  &  & 0.852 &  &  & 0.863 &  &  & 0.818 &  &  & 0.826 &  &  & 0.914 &  &  & 0.853 &  &  & 0.775 &  &  & 0.779 \\ \midrule
2              & 0.895 &  &  & 0.824 &  &  & 1.0   &  &  & 0.841 &  &  & 0.810 &  &  & 0.860 &  &  & 0.877 &  &  & 0.828 &  &  & 0.740 &  &  & 0.843 \\ \midrule
3              & 0.883 &  &  & 0.819 &  &  & 0.730 &  &  & 1.0   &  &  & 0.920 &  &  & 0.854 &  &  & 0.964 &  &  & 0.823 &  &  & 0.851 &  &  & 0.719 \\ \midrule
4              & 0.933 &  &  & 0.815 &  &  & 0.899 &  &  & 0.840 &  &  & 1.0   &  &  & 0.876 &  &  & 0.790 &  &  & 0.819 &  &  & 0.819 &  &  & 0.811 \\ \midrule
5              & 0.933 &  &  & 0.892 &  &  & 0.793 &  &  & 0.802 &  &  & 0.856 &  &  & 1.0   &  &  & 0.964 &  &  & 0.872 &  &  & 0.811 &  &  & 0.765 \\ \midrule
6              & 0.827 &  &  & 0.809 &  &  & 0.843 &  &  & 0.925 &  &  & 0.825 &  &  & 0.953 &  &  & 1.0   &  &  & 0.902 &  &  & 0.839 &  &  & 0.739 \\ \midrule
7              & 0.974 &  &  & 0.878 &  &  & 0.876 &  &  & 0.859 &  &  & 0.794 &  &  & 0.869 &  &  & 0.977 &  &  & 1.0   &  &  & 0.813 &  &  & 0.871 \\ \midrule
8              & 0.889 &  &  & 0.682 &  &  & 0.860 &  &  & 0.826 &  &  & 0.814 &  &  & 0.823 &  &  & 0.842 &  &  & 0.804 &  &  & 1.0   &  &  & 0.824 \\ \midrule
9              & 0.849 &  &  & 0.850 &  &  & 0.907 &  &  & 0.746 &  &  & 0.822 &  &  & 0.850 &  &  & 0.955 &  &  & 0.835 &  &  & 0.836 &  &  & 1.0   \\ \bottomrule
\end{tabular}
\end{table}

\begin{table}[]
\caption{Accuracy by heavily imbalanced class classification.}
\label{tab:digits-ho-ova}
\centering
\begin{tabular}{@{}cccccccccc@{}}
\toprule
\multicolumn{10}{c}{Digits   Hold-out}                                     \\ \midrule
Positive Class & Recall &  & Accuracy &  & Precision &  & F1    &  & AUC   \\ \midrule
0              & 0.955  &  & 0.206    &  & 0.107     &  & 0.192 &  & 0.539 \\ \midrule
1              & 0.972  &  & 0.218    &  & 0.112     &  & 0.201 &  & 0.552 \\ \midrule
2              & 0.858  &  & 0.212    &  & 0.098     &  & 0.176 &  & 0.500 \\ \midrule
3              & 0.803  &  & 0.188    &  & 0.093     &  & 0.167 &  & 0.460 \\ \midrule
4              & 0.994  &  & 0.176    &  & 0.108     &  & 0.195 &  & 0.539 \\ \midrule
5              & 0.950  &  & 0.249    &  & 0.114     &  & 0.204 &  & 0.560 \\ \midrule
6              & 1.0    &  & 0.154    &  & 0.106     &  & 0.192 &  & 0.529 \\ \midrule
7              & 0.843  &  & 0.346    &  & 0.116     &  & 0.204 &  & 0.567 \\ \midrule
8              & 0.965  &  & 0.204    &  & 0.105     &  & 0.190 &  & 0.544 \\ \midrule
9              & 0.966  &  & 0.193    &  & 0.107     &  & 0.193 &  & 0.536 \\ \bottomrule
\end{tabular}
\end{table}

\begin{table}[]
\caption{Resubstitution Error by balanced class classification.}
\label{tab:seme-ovo-res}
\centering
\begin{tabular}{@{}ccccccccccccccccccccccccccccc@{}}
\toprule
\multicolumn{29}{c}{Semeion Resubtitution Error}                                                                                                     \\ \midrule
\multicolumn{29}{c}{Negative Class}                                                                                                                  \\ \midrule
               & 0     &  &  & 1     &  &  & 2     &  &  & 3     &  &  & 4     &  &  & 5     &  &  & 6     &  &  & 7     &  &  & 8     &  &  & 9     \\ \midrule
Positive Class &       &  &  &       &  &  &       &  &  &       &  &  &       &  &  &       &  &  &       &  &  &       &  &  &       &  &  &       \\ \midrule
0              & 1.0   &  &  & 0.851 &  &  & 0.806 &  &  & 0.890 &  &  & 0.788 &  &  & 0.840 &  &  & 0.593 &  &  & 0.843 &  &  & 0.835 &  &  & 0.680 \\ \midrule
1              & 0.947 &  &  & 1.0   &  &  & 0.728 &  &  & 0.869 &  &  & 0.352 &  &  & 0.763 &  &  & 0.832 &  &  & 0.700 &  &  & 0.608 &  &  & 0.650 \\ \midrule
2              & 0.937 &  &  & 0.707 &  &  & 1.0   &  &  & 0.767 &  &  & 0.793 &  &  & 0.707 &  &  & 0.706 &  &  & 0.716 &  &  & 0.601 &  &  & 0.637 \\ \midrule
3              & 0.912 &  &  & 0.797 &  &  & 0.672 &  &  & 1.0   &  &  & 0.759 &  &  & 0.757 &  &  & 0.853 &  &  & 0.712 &  &  & 0.525 &  &  & 0.624 \\ \midrule
4              & 0.981 &  &  & 0.801 &  &  & 0.793 &  &  & 0.837 &  &  & 1.0   &  &  & 0.793 &  &  & 0.751 &  &  & 0.749 &  &  & 0.515 &  &  & 0.664 \\ \midrule
5              & 0.890 &  &  & 0.794 &  &  & 0.672 &  &  & 0.657 &  &  & 0.693 &  &  & 1.0   &  &  & 0.721 &  &  & 0.703 &  &  & 0.528 &  &  & 0.570 \\ \midrule
6              & 0.757 &  &  & 0.842 &  &  & 0.800 &  &  & 0.881 &  &  & 0.636 &  &  & 0.809 &  &  & 1.0   &  &  & 0.805 &  &  & 0.506 &  &  & 0.692 \\ \midrule
7              & 0.946 &  &  & 0.734 &  &  & 0.624 &  &  & 0.779 &  &  & 0.702 &  &  & 0.722 &  &  & 0.724 &  &  & 1.0   &  &  & 0.530 &  &  & 0.667 \\ \midrule
8              & 0.854 &  &  & 0.728 &  &  & 0.786 &  &  & 0.866 &  &  & 0.655 &  &  & 0.671 &  &  & 0.781 &  &  & 0.683 &  &  & 1.0   &  &  & 0.610 \\ \midrule
9              & 0.905 &  &  & 0.809 &  &  & 0.763 &  &  & 0.769 &  &  & 0.689 &  &  & 0.611 &  &  & 0.780 &  &  & 0.737 &  &  & 0.607 &  &  & 1.0   \\ \bottomrule
\end{tabular}
\end{table}

\begin{table}[]
\caption{Resubstitution Error by heavily imbalanced class classification.}
\label{tab:seme-ova-res}
\centering
\begin{tabular}{@{}cccccccccc@{}}
\toprule
\multicolumn{10}{c}{Semeion   Resubtitution Error}                         \\ \midrule
Positive Class & Recall &  & Accuracy &  & Precision &  & F1    &  & AUC   \\ \midrule
0              & 0.770  &  & 0.193    &  & 0.090     &  & 0.161 &  & 0.449 \\ \midrule
1              & 0.962  &  & 0.230    &  & 0.113     &  & 0.202 &  & 0.555 \\ \midrule
2              & 0.849  &  & 0.222    &  & 0.1       &  & 0.178 &  & 0.500 \\ \midrule
3              & 0.937  &  & 0.212    &  & 0.106     &  & 0.191 &  & 0.534 \\ \midrule
4              & 0.807  &  & 0.183    &  & 0.092     &  & 0.166 &  & 0.460 \\ \midrule
5              & 0.937  &  & 0.202    &  & 0.105     &  & 0.189 &  & 0.528 \\ \midrule
6              & 0.925  &  & 0.197    &  & 0.105     &  & 0.189 &  & 0.520 \\ \midrule
7              & 0.873  &  & 0.230    &  & 0.102     &  & 0.183 &  & 0.516 \\ \midrule
8              & 0.864  &  & 0.210    &  & 0.097     &  & 0.175 &  & 0.502 \\ \midrule
9              & 0.873  &  & 0.211    &  & 0.100     &  & 0.180 &  & 0.506 \\ \bottomrule
\end{tabular}
\end{table}

\newpage
\section{Conclusions and Future Work}
\label{section:section5}

	In this work, the performance of a fully quantum machine learning model in real datasets was tested. The evaluation is generally favorable, providing positive feedback that QML is promising. 

	Whit this work we provided a real validation to the quantum model beyond the theoretical correctness and the usual proof of concept. This showed the potential real-world application of the QML in the near future. However, it is in its early stages and in order for it to be competitive against traditional Machine Learning, there is still a gap which this work seeks to bridge. 

	Some points that would contribute to moving the QML to more mature stages are an increment in the number of available qubits to perform computation, reduce noise during the application of quantum gates, and mainly a feasible, scalable, and robust codification to map classical inputs to their quantum representation.

	As future work, we would propose to extend the biclass to multiclass classification by means of the naive extension One vs One and One vs All as a baseline. A modification in the quantum circuit generation would be proposed to allow the coding of images at three channels depth, that is, in color. It would also be interesting to implement one of the proposals for quantum convolutional layers for features extraction.

\end{document}